\documentclass[
reprint,twocolumn,
showpacs,preprintnumbers,
nofootinbib,
bibnotes,
amsmath,amssymb,
aps,
prl,
dvipdfmx
]{revtex4-1}
\usepackage{graphicx}
\usepackage{dcolumn}
\usepackage{bm}
\usepackage{comment}
\usepackage{natbib}
\usepackage{color}

\newcommand{\beq}{\begin{eqnarray}}
\newcommand{\eeq}{\end{eqnarray}}

\newcommand{\figwidth}{3.375in}
\def\vc#1{\boldsymbol #1}

\begin{document}

\title{Evidence of one-step replica symmetry breaking in a
three-dimensional Potts glass model}

\author{Takashi Takahashi}
\author{Koji Hukushima}
\affiliation{
 Graduate School of Arts and Sciences,University of Tokyo,~3-8-1,~Komaba,~Meguro-ku,~Tokyo~153-8902,~Japan
}


\date{\today}
\begin{abstract}
We study a 7-state Potts glass model in three dimensions with 
first, second, and third neighbor interactions with a bimodal distribution of couplings
by Monte Carlo simulations. Our results show the existence of a spin-glass 
transition at a finite temperature $T_c$,  a discontinuous jump of an order parameter 
at $T_c$ without latent heat, and a non-trivial structure of the order-parameter 
distribution below $T_c$. They are compatible with a one-step replica symmetry breaking. 
\end{abstract}

\maketitle


\emph{Introduction.---}
Mean-field spin-glass models without time reversal symmetry have been studied 
by many researchers over the last few decades. Being quite different from the 
Sherrington-Kirkpatrick Ising spin glass \cite{Binder1986},
a class of models such as $p$-spin model and $p$-state Potts glass model 
\cite{Crisanti1992,Gross1985}, exhibit two distinct phase transitions 
\cite{Binder2011,Castellani2005}. One is a dynamical phase transition at 
temperature $T_d$,  below which exponentially large number of metastable states 
emerge and a spin autocorrelation function does not decay to zero in the long time limit. 
The latter is a consequence of the ergodicity breaking.
The other is a purely thermodynamic transition at $T_c<T_d$,
 which is called ``random first order transition'' (RFOT).  
At $T_c$ the entropy concerned with the metastable states vanishes and 
an order parameter emerges discontinuously without latent heat, 
and below $T_c$ replica symmetry is broken at one-step level. 
A particularly intriguing fact is that at the mean-field level dynamical
equations for a time correlation function near $T_d$ in these models
are formally identical to the mode-coupling equations in the theory of 
structural glass transition. This fact suggests a potentially deep connection 
between spin-glass models with quenched disorder and structural 
glasses with no quenched disorder. 
The whole of the phenomena described above is called RFOT scenario in the field of glass transition 
and is speculated to be a promising candidate for the mean-field description
 of the glass transition.
Thus, the mean-field spin-glass theory has been developed in great	
detail, revealing that some spin-glass models 
are a prototypical model of the RFOT scenario
at least at the mean-field level \cite{Binder2011,Wolynes2012,Berthier2011}.

One of the main issues to be addressed is whether these mean-field predictions are 
valid in finite dimensions in which fluctuations must be taken into account. 
A straightforward way to investigate the effect of fluctuation is to examine finite 
dimensional spin-glass models which display RFOT in the mean-field limit. 
Previously, extensive Monte Carlo studies for the $p$-state Potts glass models 
in a three-dimensional cubic lattice clarified the existence of spin-glass transition 
at finite temperature for $p\leq 6$ \cite{lee2006,cruz2009,banos2010}.
However, their properties are rather compatible with those of continuous transition 
in the Ising spin-glass model and 
no clear remnants of RFOT have
been found. In the mean-field theory, the discontinuity of the order parameter and
also difference between $T_d$ and $T_c$ grow with the number of 
states in the $p$-state Potts glass model \cite{Caltagirone2012,Santis1995}. 
Hence, it might be likely that RFOT, if any, could be found in the Potts 
glass models with relatively large $p$ in finite dimensions. 
In addition, for such a large value of $p$, it is needed to make most of
the couplings antiferromagnetic to prevent 
ferromagnetic ordering. 
On the other hand, as pointed out  
in Ref.\cite{Cammarota2013}, 
when most of the couplings are antiferromagnetic,
the Potts glass models on any finite connectivity lattice are 
unfrustrated for large values of $p$ in a sense that these couplings are
easily satisfied in the ground state. 
Then, no glassy ordering is expected 
because the frustration is considered to be a key ingredient of the
glassy behavior. 
Indeed, Brangian \emph{et al.} found that there was no 
glassy phase in 
the $10$-state Potts glass
model with a bimodal distribution of 
the couplings with a small fraction 
of ferromagnetic couplings \cite{Brangian2003}. 
Thus, it is a difficult requirement to avoid the 
ferromagnetic ordering and to keep the frustration simultaneously
for the Potts glass models with large $p$ on finite connectivity
lattices. 
In particular, in the three dimensional Potts glass model with only
nearest-neighbor interactions, the low connectivity $c=6$ makes it
difficult to meet the requirement.

In order to avoid the above difficulties, we propose a Potts glass 
model with not only the nearest neighbor couplings,
but also second- and third-nearest neighbor couplings on a three dimensional cubic lattice.
Although this model has only short range interactions, 
such a high 
connectivity 
could yield the frustration 
even in the antiferromagnetic case and even for large $p$. 
Using Monte Carlo simulations for the $p$-state Potts glass model with
$p=7$, we obtained the following results: 
(1) This model shows a static spin-glass transition at finite temperature
$T_c/J=0.421(3)$ with the correlation length exponent $\nu =0.68(9)$. 
(2) At $T_c$ the order parameter appears discontinuously but no latent heat exists.
(3) Below $T_c$ the order-parameter distribution has a bimodal structure. 


\emph{Model and Numerical Details. ---}
The $p$-state Potts glass model we studied is defined by the Hamiltonian 
\begin{eqnarray}
	\mathcal{H}_{\mathcal{J}}(\sigma)=-\sum_{(i,j)}J_{ij}\delta(\sigma_i,\sigma_j)
	\label{eq:hamiltonian},
\end{eqnarray}
where the Potts spin $\sigma_i$ on the site $i$ takes $0,1,\ldots,p-1$ and the summation is over
the nearest, second-nearest and third-nearest neighbors on a
three-dimensional cubic lattice of size $N=L^3$ with periodic
boundaries. 
Each of the sites has connectivity $c=26$, and 
a set of coupling constants $\mathcal{J}=\left\{J_{ij} \right\}$ are quenched
random variables chosen from a bimodal distribution
$P(J_{ij})=x\delta(J_{ij}-J)+(1-x)\delta(J_{ij}+J)$, where $x$ denotes
the fraction of ferromagnetic couplings. 
In order to prevent a ferromagnetic transition, we set
$x=(1-1/\sqrt{2})/2\simeq 0.15$ and $J=\sqrt{2}J_0$. Then, the mean and
variance of the couplings are $-1$ and $1$, respectively, measured in
the unit of $J_0$. 
This means that 
most of the couplings are antiferromagnetic in this model. 
Note that because all the spins in the smallest cube with $L=2$ 
interact with each other through up to the third-neighbor couplings, 
there remains the finite frustration for $p\leq 7$ even in the purely antiferromagnetic case. 
In this Letter, we focus on the case of $p=7$.

\begin{figure}[b]
	\includegraphics[width=\figwidth]{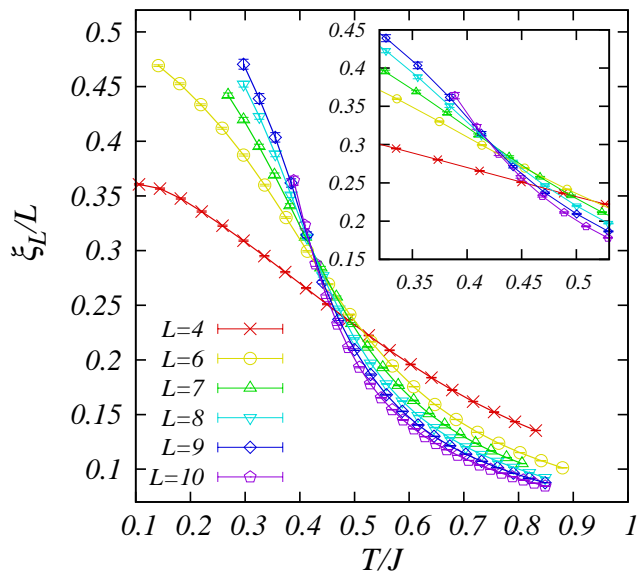} 
	\caption{\label{fig:xiloverl}(Color Online) Temperature dependence of the dimensionless correlation length $\xi_L/L$. The inset shows its enlarged view around the transition temperature.
}
\end{figure}

Since spin-glass simulations are hampered by extremely slow relaxation
dynamics, we use replica exchange Monte Carlo
method \cite{hukushima1996}.  
The linear sizes are $L=4-10$ for most of observables explained 
below and $L=14$ for the energy density and the specific heat which are
relatively easy to evaluate. 
The number of samples averaged over is $256-4096$  depending on the
system size. 
The total number of Monte Carlo sweeps (MCS) used on each lattice size is
$10^6-10^8$. 
We examined equilibration 
by monitoring the Monte Carlo average of the observables 
while doubling the number of MCS for measurement successively.
The data are regarded as equilibrium values when the last two data agree
within their error bars. 


\emph{Observables.-- }It is convenient 
to represent the Potts variables using 
the simplex representation~ \cite{ZiaWallace},  in which the spin
variable $\vc{S}_i$ of the site $i$  takes one of $p$ unit
vectors $\{ \vc{e}^{(\alpha)}\}_{\alpha=1}^p$ 
pointing to the corner of the simplex in the  $p-1$ dimensional space. 
These vectors satisfy the relations
$\vc{e^{(\alpha)}}\cdot\vc{e^{(\beta)}}={(p\delta_{\alpha,\beta}-1)}/{(p-1)}$. Some
observables calculated in our simulations are 
expressed as those in vector spin glasses 
using the simplex representation. 
To study the spin-glass transition we define a spin-glass order parameter as an overlap
between two replicas.
For two independent replica configurations denoted as
$\{\vc{S}_i^{(1)}\}_{i=1}^N$ and $\{\vc{S}_i^{(2)}\}_{i=1}^N$ with the same disorder, 
the wave-number dependent overlap between them 
for the Potts-glass model is defined by a tensor $q^{ab}(\vc{k})$: 
\begin{eqnarray}
	q^{ab}(\vc{k})=\frac{1}{N}\sum_{i=1}^N S_i^{a,(1)}S_i^{b,(2)}e^{i\vc{k}\cdot\vc{R}_{i}},
	\label{eq:qab}
\end{eqnarray}
where the upper suffixes $a$ and $b$ are indices of the simplex vector
component and $\vc{R}_{i}$ is a displacement vector at the site $i$. 
A rotational invariant scalar overlap is also defined by 
\begin{eqnarray}
	q(\vc{k})=\sqrt{\sum_{a,b}^{p-1}\left|q^{ab}(\vc{k})\right|^2}
	\label{eq:overlap}.
\end{eqnarray}
Then, the wave-number-dependent spin-glass susceptibility
$\chi_{\mathrm{SG}}(\vc{k})$ is given by an expectation value 
\begin{eqnarray}
	\chi_{\mathrm{SG}}(\vc{k})=
		N\left[\langle q^2(\vc{k})\rangle^{(T)}\right]_{\rm av},
	\label{eq:chisg}
\end{eqnarray}
where $[\cdots]_{\rm av}$ and $\langle\cdots\rangle^{(T)}$ represent
an average over the quenched disorder and a thermal average at 
temperature $T$, respectively. 
The dimensionless correlation length $\xi_L/L$ is useful for estimating
the critical temperature $T_c$ because it is independent of $L$ at
$T_c$. Thus, the intersection temperature in the plot of $\xi_L/L$ for
various $L$ gives the estimate of $T_c$. 
The finite-size correlation length $\xi_L$ is estimated from
$\chi_{\mathrm{SG}}(\vc{k})$ as  \cite{ballesteros2003}
\begin{eqnarray}
	\xi_L=\frac{1}{2\sin\left(|\vc{k}_{\mathrm{min}}|/2\right)}\sqrt{\frac{\chi_{\mathrm{SG}}(\vc{0})}{\chi_{\mathrm{SG}}(\vc{k_{\mathrm{min}}})}-1}
	\label{eq:finitexi},
\end{eqnarray}
where $\vc{k_{\mathrm{min}}}=(2\pi/L,0,0)$ is the smallest nonzero wave
vector.
Another dimensionless quantity is the Binder parameter defined by
\begin{eqnarray}
	g_4=\frac{(p-1)^2}{2}\left(1+\frac{2}{(p-1)^2}-\frac{\left[\langle
							      q^4(\vc{0})\rangle^{(T)}\right]_{\rm av}}{\left[\left\langle q^2(\vc{0})\right\rangle^{(T)}\right]_{\rm av}^2}\right). 
	\label{binder}
\end{eqnarray}
This quantity is known to exhibit a peculiar behavior for systems
with a one-step RSB (1RSB) transition \cite{Picco2001,HK2000,Billoire2005},
while it is expected to exhibit the intersection at a conventional second
order transition temperature. 

One of the most important quantities for studying  
the phase space structure of the spin-glass phase 
is the overlap distribution function
\begin{eqnarray}
	P^{(T)}(Q)=\left[\left\langle
				\delta\left(Q-q(\vc{0})\right)\right\rangle^{(T)}\right]_{\rm
	av},
	\label{eq:pofq}
\end{eqnarray}
which is accessible from Monte Carlo simulations. 
The overlap distribution function has a non-trivial structure if the
replica symmetry breaking occurs. In particular, two separated peaks
appear in $P^{(T)}(Q)$ at and below $T_c$ for a 1RSB system, that is similar
to the order-parameter distribution found in systems with a first-order
transition.


\begin{figure}[b]
	\includegraphics[width=\figwidth]{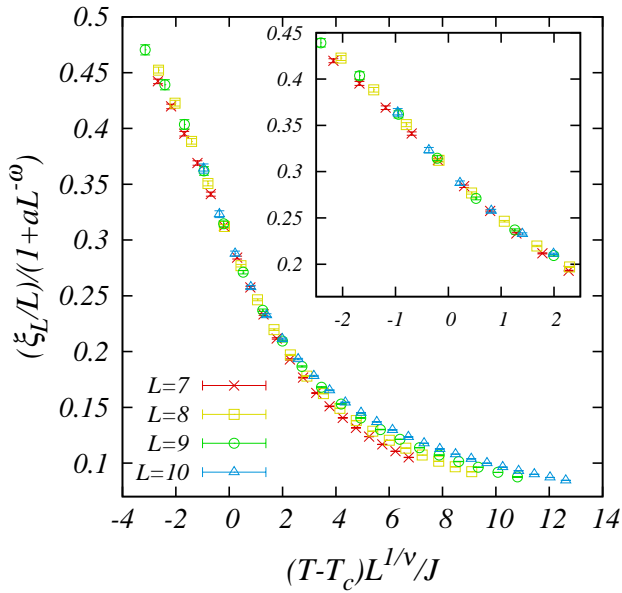}
	\caption{\label{fig:xiscaling} (Color Online) Scaling plot of the
 finite-size correlation length ratio $\xi_L/L$ according to
 Eq.(\ref{eq:xiscaling}) using $T_c/J=0.421, \nu=0.68, a=1$ and $\omega=3$. The inset shows its magnified view.}
\end{figure}

\begin{figure}[t]
	\begin{tabular}{l}
		\includegraphics[width=\figwidth]{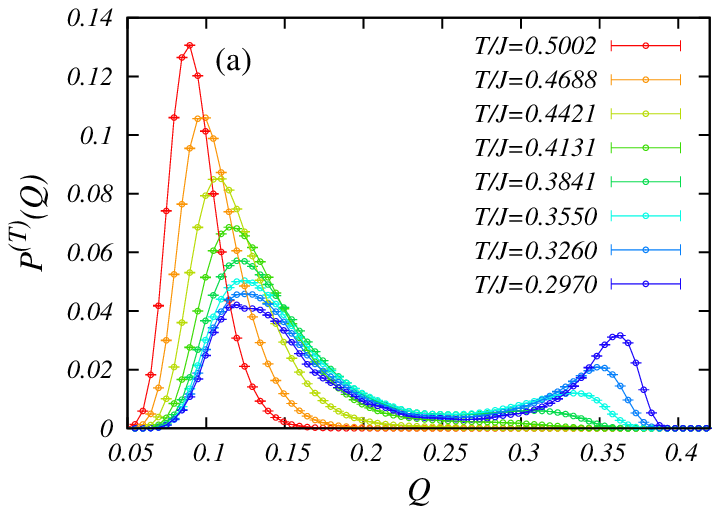}\\
		\includegraphics[width=\figwidth]{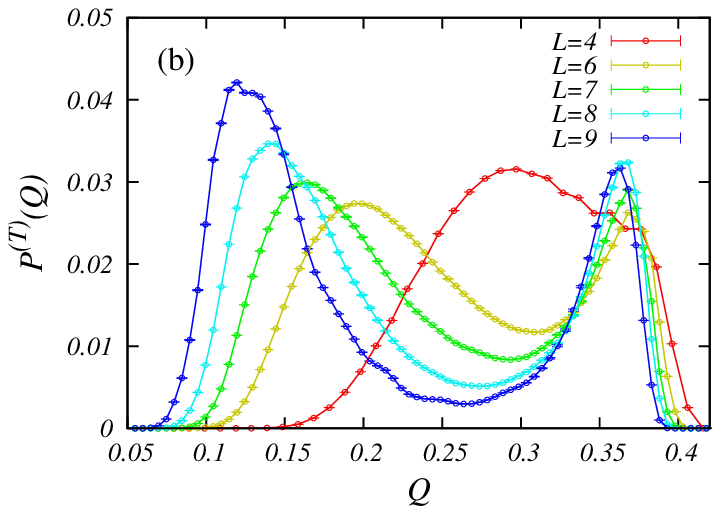}
	\end{tabular}
	\caption{\label{fig:pofq}(Color Online) Overlap distribution
 function of a Potts glass model in three dimensions for various
 temperatures with $L=9$ (a) and for different sizes at $T/J=0.2970$
 below $T_c$ (b). 
}
\end{figure}

\emph{Numerical results.---}
First, in order to investigate critical properties of the Potts glass
model, we see 
the finite-size correlation length $\xi_L$ scaled by $L$.  
As shown in Fig.~\ref{fig:xiloverl}, 
a clear intersection is observed around $T/J\simeq 0.4$, though it is
slightly shifted to low temperature with increasing $L$. 
The intersection for asymptotically large $L$ provides an evidence of
spin-glass phase transition at the temperature. 
The Potts glass model for $p=7$ with nearest neighbor interactions has
no glassy phase at up to very low temperature, possibly down to zero
with the present fraction of ferromagnetic couplings. 
The second- and third-neighbor couplings cause the spin-glass transition temperature
to increase significantly. 
To determine $T_c$ and $\nu$, 
we perform 
a finite-size scaling analysis in which the dimensionless correlation
length is assumed to follow the scaling form,  up to the leading
correction term,  
\begin{eqnarray}
	\frac{\xi_L}{L}=\widetilde{X}\left((T-T_c)L^{1/\nu}\right)(1+aL^{-\omega}),
	\label{eq:xiscaling}
\end{eqnarray}
where $\nu$ is the correlation length exponent, $\omega$ is an
exponent of the leading correction,  
and $\widetilde{X}$ is an universal scaling function. 
The scaling parameters such as $T_c$ and $\nu$ are determined 
by requiring all the curves of $\xi_L/\left(L(1+aL^{-\omega})\right)$
against $(T-T_c)L^{1/\nu}$ to collapse on a 
single curve near $T_c$. 
Bayesian scaling analysis recently developed \cite{harada2011} is used 
to perform the scaling analysis systematically. 
Fig.\ref{fig:xiscaling} shows the scaling plot of 
$\xi_L/L$, which is obtained by 
\begin{eqnarray}
	T_c/J=0.421(3)~,~\nu=0.68(9)
	\label{eq:critical}.
\end{eqnarray}
The value of $\nu$  is consistent with $2/d$ where $d$ is the spatial
dimension, derived by a heuristic scaling argument based on
RFOT \cite{Kirkpatrick2014}, suggesting that the overlap function has a
finite jump at $T_c$. This is in contrast with the fact that the value of
$\nu$ is slightly larger than $2/d$ in the three-dimensional Potts-glass
models with the nearest-neighbor couplings \cite{lee2006,cruz2009,banos2010}.

In Fig.~\ref{fig:pofq}, temperature  and system-size dependence of
$P^{(T)}(Q)$ is shown. At high temperatures the distribution function
has a single Gaussian-like peak near $Q\simeq 0$.
The peak position is expected to approach zero in the thermodynamic
limit. 
On the other hand, below $T_c$ another peak at a larger value of $Q$,
corresponding to the Edwards-Anderson order parameter $q_{\mathrm{EA}}$, 
emerges and coexists with the other peak at lower $Q$. 
The lower panel in Fig.~\ref{fig:pofq} shows the size dependence of
$P^{(T)}(Q)$ at $T/J=0.2970$, which is well below the estimated $T_c$. 
The peaks at $Q=q_{EA}$ and $Q\simeq0$ show a tendency to 
grow in height and become narrower in width with increasing $L$. 
Further, the weight between these two peaks is strongly suppressed with $L$. 
These imply that the bimodal structure in $P^{(T)}(Q)$ remains in the
thermodynamic limit, providing 
a clear evidence of the 1RSB nature in the spin-glass phase.

\begin{figure}[t]
		\includegraphics[width=7.0cm]{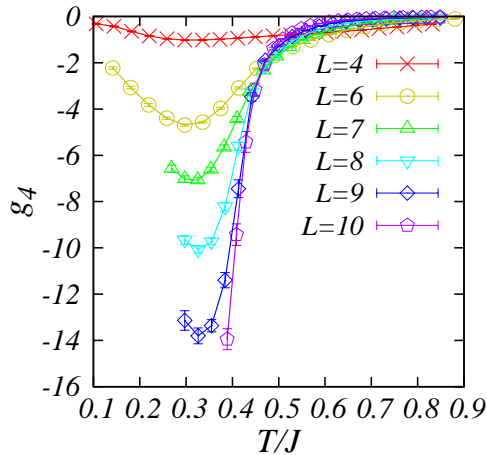} 
	\caption{\label{fig:binder}(Color Online) 
Temperature dependence of the Binder parameter $g_4$ of a Potts glass
 model in three dimensions. 
}
\end{figure}

While Fig.~\ref{fig:pofq}(a) suggests that the overlap emerges
discontinuously at $T_c$, 
the peak of $P^{(T)}(Q)$ near $T_c$ is rounded by the finite-size effect. 
Another evidence of the discontinuous jump is
found, however,  in temperature dependence of the Binder parameter $g_4$. 
As shown in Fig.~\ref{fig:binder}, $g_4$ 
exhibits a negative dip near $T_c$ with a negatively divergent
tendency for large $L$. 
Note that $g_4\to -\infty$ at $T_c$ when a 1RSB transition with a finite jump of $q_{\mathrm{EA}}$ occurs in the mean-field glass models \cite{Picco2001,Billoire2005}, in contrast to a
continuous full RSB transition and also an ordinary second order phase transition. Thus,  this divergent behavior
implies that $q_{\mathrm{EA}}$ 
appears at $T_c$ discontinuously.

\begin{figure}[t]
	\begin{tabular}{ll}
		\hspace{-0.47cm}
		\includegraphics[width=4.8cm]{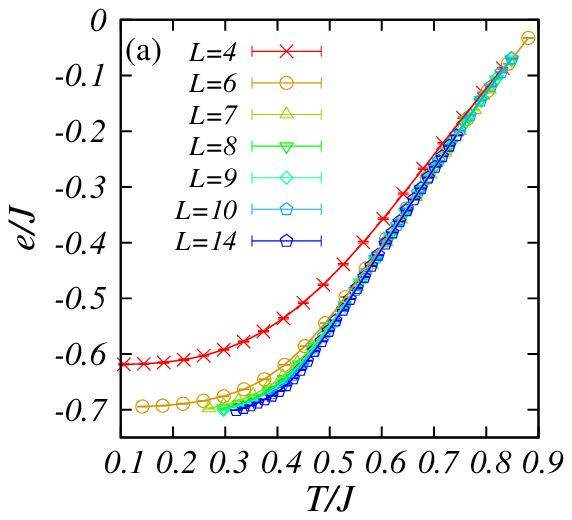}&
		\hspace{-0.65cm}
		\includegraphics[width=4.8cm]{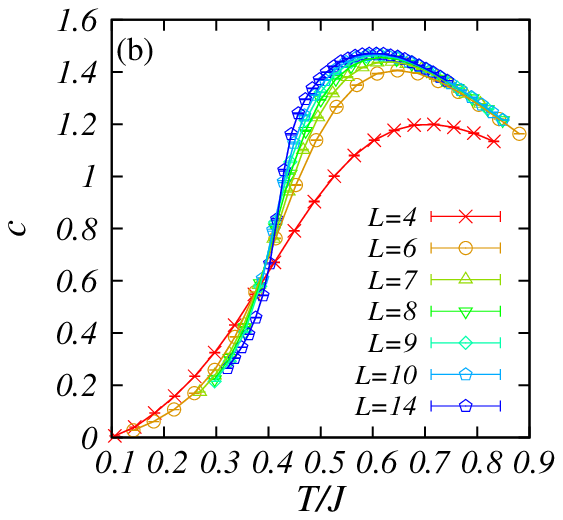}
	\end{tabular}
	\caption{\label{fig:energy}(Color Online) 	
Temperature dependence of the energy density (a) and the specific heat
 (b) of a Potts glass model in three dimensions. 
}
\end{figure}

Finally, Fig.~\ref{fig:energy} shows temperature dependence of the
energy density and the specific heat. 
No discontinuity of the energy density, and hence no divergent tendency in
the specific heat are observed at around $T_c$. 
Instead, the specific heat for various sizes has an
intersection near $T_c$.  
This might indicate that in the thermodynamic limit the specific heat has 
a discontinuous jump 
at $T_c$ as expected from some mean-field spin-glass models
with RFOT. Further study is required to clarify this point.

\emph{Conclusions.---}
In this letter, the $7$-state Potts glass model with the nearest, second-nearest and third-nearest neighbor 
interactions has been proposed
as a candidate for displaying RFOT in finite dimensions. 
A key ingredient is to keep both a large number of Potts states and
the frustration. 
All of our equilibrium numerical 
results suggest that the present model in three dimensions 
shares many features of RFOT, namely 
a spin-glass transition at finite temperature, 
a jump of the spin-glass order parameter at $T_c$ 
without latent heat, and 
a bimodal overlap distribution below $T_c$, as expected
from 1RSB. 
Thus, 
we conclude that this is the first realization of the finite-dimensional
statistical-mechanical model which mimics a static part of the whole
RFOT scenario. 
Another important aspect of the RFOT scenario is dynamical properties, 
which are believed to be modified in finite dimensions from the mean-field predictions.
This model provides a promising test bed for further examining the validity of the
RFOT scenario in finite dimensions,  
which remains to be investigated.

\begin{acknowledgments}
The authors thank S.~Sasa for useful discussions. 
The authors also thank S. Takabe and Y. Nishikawa for carefully reading the manuscript.
Numerical simulation in 
this letter has mainly been
performed by using the facility of the Supercomputer Center, Institute
for Solid State Physics, University of Tokyo. 
This research was supported by the Grants-in-Aid for Scientific Research 
from the 
{JSPS}, 
Japan (Nos. 22340109, 25120010  and 25610102), and 
JSPS Core-to-Core program ``Nonequilibrium dynamics of soft matter and information.''
\end{acknowledgments}

\end{document}